\documentclass[sigconf]{acmart}

\usepackage{booktabs}	
\usepackage{diagbox}    
\usepackage{multirow}   
\usepackage{verbatim}	
\usepackage{array}    
\usepackage{balance} 

\usepackage{amsmath,amssymb,amsfonts,graphicx}
\usepackage{epstopdf}

\copyrightyear{2023}
\acmYear{2023}
\setcopyright{acmlicensed}\acmConference[WWW '23 Companion]{Companion Proceedings of the ACM Web Conference 2023}{April 30-May 4, 2023}{Austin, TX, USA}
\acmBooktitle{Companion Proceedings of the ACM Web Conference 2023 (WWW '23 Companion), April 30-May 4, 2023, Austin, TX, USA}
\acmPrice{15.00}
\acmDOI{10.1145/3543873.3587584}
\acmISBN{978-1-4503-9419-2/23/04}

\begin{document}

\title{Federated Learning for Metaverse: A Survey}

\author{Yao Chen}
\authornote{Both authors contributed equally to this work.}
\affiliation{%
	\institution{Jinan University} 
	\city{Guangzhou}
	\country{China}
}
\email{csyaochen@gmail.com}

\author{Shan Huang}
\authornotemark[1]
\affiliation{%
	\institution{Jinan University} 
	\city{Guangzhou}
	\country{China}
}
\email{shuang9901@gmail.com}

\author{Wensheng Gan}
\authornote{Corresponding author, also with Pazhou Lab, Guangzhou 510330, China}
\affiliation{%
	\institution{Jinan University} 
	\city{Guangzhou}
	\country{China}
}
\email{wsgan001@gmail.com}

\author{Gengsen Huang}
\affiliation{%
	\institution{Jinan University} 
	\city{Guangzhou}
	\country{China}
}
\email{hgengsen@gmail.com}

\author{Yongdong Wu}
\affiliation{%
	\institution{Jinan University} 
	\city{Guangzhou}
	\country{China}
}
\email{wuyd007@qq.com}

\renewcommand{\shortauthors}{Chen et al.}

\begin{abstract}
   The metaverse, which is at the stage of innovation and exploration, faces the dilemma of data collection and the problem of private data leakage in the process of development. This can seriously hinder the widespread deployment of the metaverse. Fortunately, federated learning (FL) is a solution to the above problems. FL is a distributed machine learning paradigm with privacy-preserving features designed for a large number of edge devices. Federated learning for metaverse (FL4M) will be a powerful tool. Because FL allows edge devices to participate in training tasks locally using their own data, computational power, and model-building capabilities. Applying FL to the metaverse not only protects the data privacy of participants but also reduces the need for high computing power and high memory on servers. Until now, there have been many studies about FL and the metaverse, respectively. In this paper, we review some of the early advances of FL4M, which will be a research direction with unlimited development potential. We first introduce the concepts of metaverse and FL, respectively. Besides, we discuss the convergence of key metaverse technologies and FL in detail, such as big data, communication technology, the Internet of Things, edge computing, blockchain, and extended reality. Finally, we discuss some key challenges and promising directions of FL4M in detail. In summary, we hope that our up-to-date brief survey can help people better understand FL4M and build a fair, open, and secure metaverse.
\end{abstract}



\maketitle

\section{Introduction}

The meaning of the word "metaverse" \cite{sun2022metaverse,sun2022big,chen2022metaverse} is a world beyond reality. At the technical level, the metaverse can be regarded as the integration mechanism or carrier of big data and information technology \cite{wang2022survey}. The metaverse combines various technologies and hardware, such as virtual reality, digital twins, the Internet of Things, blockchain, digital collections, and so on \cite{kang2022blockchain}. The main goal of the metaverse is to build a new world of interaction between the real and the virtual. All things in the real world can be digitally replicated in the metaverse and interact with and influence the real world \cite{wangyitong2022survey}. The metaverse is the human imagination of digital life. It also inspires the imagination about the future development of the Internet. From the perspective of digital civilization, the metaverse will drive the development of digital civilization.

\begin{figure}[h]
    \center
    \includegraphics[clip,scale=0.2]{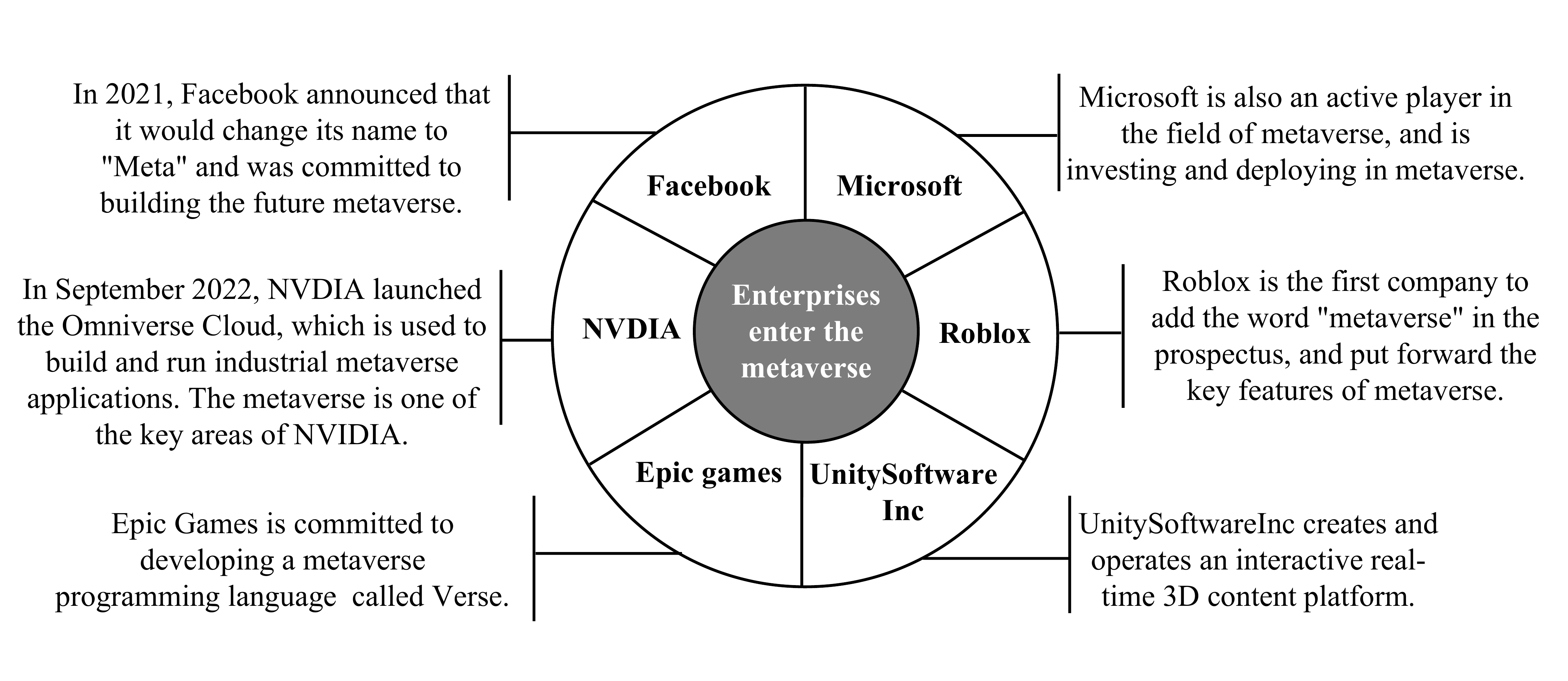}
    \caption{Enterprises enter the metaverse.} 
    \label{Meta}
\end{figure}

In recent years, driven by practical needs and the feasibility of the construction of the metaverse, the metaverse has attracted worldwide attention. The metaverse has aroused much discussion and research in various fields. Many technology companies are also attracted by the metaverse \cite{pham2022artificial}. Figure \ref{Meta} lists some companies that have invested in the metaverse field. Analyze the reasons why the metaverse is important from the perspective of practical needs. People want to work more efficiently under any circumstances and want social and entertainment methods to become more diverse. From the perspective of construction feasibility, the rapid developments of the Internet of Things, artificial intelligence, virtual reality, edge computing, blockchain, and other technologies provide the feasibility of the metaverse's construction \cite{yang2022fusing}.

At present, the metaverse is in a stage of innovation and exploration, and it is full of chaos in terms of privacy data disclosure, which needs to be regulated and managed. The metaverse can be seen as the integration mechanism or carrier of big data and information technology. In general, the metaverse needs a lot of data for support. However, these data are often difficult to obtain. On the one hand, data privacy and security issues have attracted much attention. Governments have enacted many laws to protect user privacy, such as the GDPR \cite{voigt2017eu} issued by the European Union and the CCPA \cite{pardau2018california} issued by the United States. On the other hand, there is competition in the industry. Sharing data may harm the company's own interests. There is no doubt that the metaverse is facing the challenges of data privacy and security and the difficulties of data collection at the data level.

Federated learning (FL) \cite{yang2019federated} is a powerful technique that can solve the metaverse dilemma at the data level. FL is a distributed machine learning paradigm with a privacy protection function, designed for a large number of edge devices \cite{kairouz2021advances}. FL allows the original data to be retained on each edge device and uses the data, computing power, and model-building capabilities of each edge device to perform tasks \cite{zeng2021fedlab,chen2023privacy}. The server in FL collects and aggregates the parameters or models sent by the edge devices. The operation of FL training models is an iterative process of model distribution, collection, aggregation, and update \cite{chen2022federated}. In FL, each round of iteration can be divided into three steps. First, the central server generates a global model using the existing data or updates the global model using the collected parameters. Then, this global model is sent to all participants. Each participant trains this global model using the local data to obtain some parameters. Finally, the participants send the parameters to the central server. The central server uses these parameters to update the global model. Until the model reaches convergence, FL stops training the model. In order to further improve the data privacy protection ability of the metaverse and help collect data, this paper discusses the integration of the metaverse and FL.

Federated learning for the metaverse (FL4M) is a promising research direction for the following reasons. First, FL ensures that the original data does not leave the user. This is because the user only needs to transmit parameters or models to third parties. It guarantees the privacy and security of the user's data. Therefore, combining FL into the metaverse improves data security. Secondly, FL is a distributed machine learning method. FL trains the model by using each user's data, computing ability, and model building ability. Therefore, combining FL into the metaverse reduces the demands on the memory and computing capacity of the central server. In addition, FL can introduce contribution measurement mechanisms and incentive strategies. They come into play when dealing with cross-silos settings. FL evaluates the contribution value of each user based on the contribution measurement mechanism and provides incentives. In this way, more users will be attracted to participate in the model training. Therefore, integrating FL's contribution measurement mechanism and incentive strategy into the metaverse will increase the interest of users and the accuracy of the model.

\textbf{Contributions}: The research in this paper focuses on the advances and opportunities presented by FL4M. The main contributions of this paper are as follows.

\begin{itemize}
    \item To the best of our knowledge, this is the first paper discussing the advances and opportunities of FL for the metaverse. We explain why the FL should be integrated into the metaverse. In many scenarios, FL has the advantage of effectively helping the metaverse. Besides, we analyze the reasons why FL4M is a promising research direction.

    \item  We discuss for the first time the convergence of key metaverse technologies and FL, such as big data, communication technologies, the Internet of Things, edge computing, blockchain, and extended reality.

    \item  We highlight some key challenges and promising directions in FL4M and try to put forward some enlightening ideas for this research direction. In addition, our survey aims to help researchers understand the up-to-date development of FL4M. We hope that our survey can help researchers understand and establish a fair, open, and safe metaverse.
\end{itemize}

\textbf{Organization}: The rest of the paper is organized as follows. In Section \ref{sec:relatedwork}, we discuss related concepts, including metaverse, artificial intelligence, FL, and artificial intelligence in the metaverse. Then, in Section \ref{sec:keytechnologies}, we describe FL for key metaverse technologies in detail. Section \ref{challenges} and Section \ref{directions} respectively discuss the challenges and promising directions of FL4M. Finally, Section \ref{sec:conclusion} summarizes this paper. The important acronyms are listed in Table \ref{abbreviations}.

\begin{table}[!h]
	\centering
	\caption{Summary of important abbreviations}
	\label{abbreviations} 
	\begin{tabular}{cc}
		\hline
		Abbr. &   Definition \\ \hline
		FL    & Federated Learning    \\
		AI    & Artificial Intelligence  \\
		NNs    & Neural Networks   \\
		DL    & Deep Learning \\
        ML  & Machine Learning \\
        IoT  & Internet of Things \\
        MEC  & Mobile Edge Computing \\
        XR  & Extended Reality \\
		VR    & Virtual Reality \\ 
        AR  & Augmented Reality \\
		MR    & Mixed Reality \\
        HCI  & Human-Computer Interaction \\
		\hline
		\\
	\end{tabular}
\end{table}

\section{Related Concepts} \label{sec:relatedwork}

\subsection{Metaverse}

The term "metaverse" originated from the American science fiction novel \textit{Snow Crash} \cite{joshua2017information, lee2021all}, which constructed a virtual space parallel to and interacting with real space. Unlike mobile terminals such as cell phones and computers, which are only in flat space, the metaverse is a three-dimensional space-time Internet \cite{ning2021survey}. By merging network technology and extended reality, the metaverse achieves the goal of integrating virtual life with natural life and machine life \cite {owens2011empirical}. 2021 is the first year of the rapid development of the metaverse, and academic and media circles all have deep expectations for the realization of the metaverse \cite{sun2022metaverse}. In fact, the metaverse is not a new concept or technology but an integration of existing technologies, linking networks, hardware terminals, and users, thus forming a virtual display system that is closely connected to the real world yet highly independent \cite{mystakidis2022metaverse}. Key technologies in the metaverse include data science \cite{gan2019survey,gan2021survey}, big data \cite{gan2017data,sun2022big}, Internet of behavior \cite{sun2023internet}, artificial intelligence \cite{zhang2022artificial}, interactivity \cite{zauskova2022visual}, communication technology \cite{yu20226g}, edge computing \cite{siriwardhana2021survey}, IoT \cite{shafique2020internet}, blockchain \cite{ryskeldiev2018distributed}, and networking \cite{tang2022roadmap}. Better than the traditional Internet, the realistic simulated world created by the metaverse provides a more real-time and immersive experience for users. The realization of the metaverse requires three steps, including digital twins \cite{liu2021review}, virtual natives, and virtual-real fusion \cite{wang2022survey}. The term "digital twins" refers to the mapping of the real world by replicating it in the metaverse. For example, in Snow Crash, each user has incarnations in the metaverse, i.e., a precise copy is created in the virtual world. Virtual world scenes and users are also able to interact like in the real world, thus achieving "virtual natives" status. The third realm is an integrated symbiotic system where virtual and reality interact and influence each other.

\subsection{Artificial Intelligence}

Artificial intelligence (AI) \cite{kotsiantis2007supervised, russell2010artificial} is intelligence exhibited by machines with the ability to perceive, synthesize, and infer information. For example, the application scenarios of AI in pattern recognition \cite{bishop2006pattern} include computer vision \cite{voulodimos2018deep}, machine translation \cite{lu2019artificial}, etc. In addition, the cognitive functions of AI can be used for predictive judgment, such as in autonomous driving \cite{zhang2022artificial}, AlphaGo \cite{strong2016applications}, etc. AI maximizes efficient and successful responses or decisions by sensing the external environment and imitating the way humans think. AI is divided into two main categories: weak artificial intelligence and strong artificial intelligence. Weak artificial intelligence refers to AI that only focuses on one area without an independent will, such as voice assistants and face recognition. Strong artificial intelligence, on the other hand, is capable of acting autonomously outside of a set program and can even surpass humans in some respects, such as with driverless cars and intelligent robots. Research areas of AI include machine learning (ML) \cite{mitchell2007machine}, artificial neural networks (NNs) \cite{yegnanarayana2009artificial}, deep learning (DL) \cite{lecun2015deep}, etc. Among them, machine learning, as a method to achieve artificial intelligence, generally employs algorithms to parse and learn from data to accomplish decision-making or prediction tasks. As an implementation technique in machine learning, deep learning has enabled greater breakthroughs in machine learning with applications in a wider range of fields \cite{lu2019artificial}. It is mentioned in \cite{pham2022artificial} that AI can provide product inspection and fault diagnosis in the metaverse.  A number of CNN-based and RNN-based frameworks have been proposed to build intelligent fault detection systems in manufacturing. Examples include the data-driven learning model on CNN architecture with a transfer learning mechanism \cite{guo2018deep} and the deep encoder-decoder framework on RNN architecture with an attention mechanism \cite{lee2020attention}.

\subsection{Federated learning}

Federated learning was originally proposed by Google to update models locally for Android phones \cite{mcmahan2017communication}. The purpose of federated learning is to carry out efficient ML while safeguarding the privacy of the user and terminal data \cite{yang2019federated}. Thus, essentially, federated learning is a distributed machine learning technique \cite{konevcny2016federated} used to provide data privacy protection. Depending on the distribution of data among multiple participants, federated learning is divided into three main categories: horizontal federated learning, vertical federated learning, and federated transfer learning \cite{yang2019federated}.

$\bullet$ In this case, horizontal federated learning is to distribute the sample data to each machine, and then these machines download the model from the server separately for training, and finally, the parameters will be fed back to each machine. This type of federated learning is suitable for scenarios where the samples are different, but the features between the samples are similar. 

$\bullet$  The applicable scenario of vertical federated learning is the opposite of horizontal federated learning, i.e., the union of features of cross-users under different businesses. During the learning process, each participant is unaware of the existence of other parties and only uploads their model parameters, but the prediction requires joint modeling and collaboration to complete the task. 
     
$\bullet$  Federated transfer learning can be considered useful when there is no significant overlap between either the samples or their features. The key problem is to identify similarities between the source and target domains. 
     
Federated learning \cite{li2020federated} takes the distributed data and trains the model locally for data security. With a trusted channel established by secure algorithms such as homomorphic encryption, the results of the model can be aggregated to a central node for training to obtain the final training model. Federated learning ensures both parity and independence among the participants. Moreover, the entire process of learning and training achieves data isolation to prevent data leakage, which well protects the privacy of users \cite{konevcny2016federated}. In such a distributed machine learning framework, the various participants can be coordinated to build more accurate global models. Until now, there are many studies about federated learning.

\subsection{Artificial Intelligence in the Metaverse}

Artificial intelligence is an essential technological support for the realization of the metaverse platform. On the one hand, the real world is mapped to the virtual world through AI technologies such as visual recognition. On the other hand, with the help of AI algorithms such as image generation, the virtual world is also able to influence the real world. The essence of the metaverse is the combination of various technologies to create a virtual world that interacts with reality. Therefore, the combination of AI with VR, AR, and other technologies is important for creating a reliable and efficient metaverse. AI technology can not only improve the performance of the facilities in the metaverse but also provide security for their infrastructure \cite{pham2022artificial, zhu2022metaaid}. For example, the algorithm proposed by Yampolskiy \textit{et al.} \cite{yampolskiy2012face} is able to accurately identify and verify virtual characters. In the metaverse, law enforcement agencies also need this type of technology to track a character who may have committed a criminal act \cite{yang2022fusing}. In addition, the metaverse needs to be supported by communication technologies, and there are already many AI technologies \cite{chen2019artificial, she2020deep, alsenwi2021intelligent} that are widely used in network architecture applications that provide reliable and low-latency network access and security.

\section{Key FL4M Technologies} \label{sec:keytechnologies}

As shown in Figure \ref{Key technologies}, FL4M mainly has the following six key techniques. The combination of FL and big data, communication technology, IoT, edge computing, blockchain, and extended reality technology can help alleviate the privacy leakage problem in the metaverse and facilitate the sharing of data resources.

\subsection{Big Data of FL4M}
Big data is the foundation for metaverse applications.

$\bullet$  \textbf{Purposes}. Big data refers to data that contains many types and whose volume is large and requires fast processing speed. Therefore, the acquisition, storage, analysis, and protection of big data is a crucial and challenging task. In the metaverse, the constructed digital space is essentially data. In other words, the metaverse can be regarded as a fusion mechanism for big data and various information technologies. When building a bridge between virtual and reality, real-time data gathering and processing capabilities of sensors and smart devices are the cornerstones for supporting large numbers of users online at once. 

$\bullet$  \textbf{Challenges}. In the midst of increased privacy protection and data oversight, the phenomenon of data existing as isolated islands has emerged \cite{yang2019federated}. The rich data of multiple business departments in an enterprise is defined by each other independently. Data from a single department is similar to an isolated island that is incapable of interacting with other rich data \cite{li2020federated}. In the era of the metaverse, data is undoubtedly one of the most important assets of an enterprise. Therefore, it is vital to shatter the isolated island of data in order to actualize information exchange and the hidden value underlying big data \cite{sun2022big}.

$\bullet$  \textbf{Methods}. In compliance with data security management, each participant in the federated model is able to use its own data for joint training with others in a safe and fair manner without causing data leakage. In the metaverse, each user can be abstracted as a data owner. Utilizing all the potential value that may be found in data and fully optimizing business technology and services are both possible with the help of big data and FL technology.

$\bullet$  \textbf{Applications}. FL, a new generation of technology, is applied in finance, healthcare, and security to address isolated data islands and privacy protection issues \cite{li2021survey}. FATE is the first industrial-level federated learning open-source framework developed by WeBank, which can build federated models of general business scenarios.

\subsection{Communication Technology of FL4M}

$\bullet$  \textbf{Purposes}. It can be expected that the metaverse will generate huge data throughput. When it comes to human-computer interaction, a huge user community demands lower latency. Therefore, the interaction between the virtual and real worlds in the metaverse is based on high-performance communication technologies. Wired and wireless communication are the two main categories of communication. Without taking into account the distance between computers and communication devices, wireless communication is more flexible, making it more appropriate for the metaverse.

$\bullet$  \textbf{Challenges}. Wireless communication technology has evolved to 5G at present. However, its transmission speed is still some way from the communication goal of the metaverse. In this case, future 6G wireless networks need to address high throughput, low latency, and the security of data transmission in the metaverse.

$\bullet$  \textbf{Methods}. Yang \textit{et al.} \cite{yang2022federated} proposed the use of federated learning for 6G networks. Federated learning is able to conserve wireless resources and reduce transmission latency with distributed management. Moreover, sharing predictive models and collaborative learning are needed for federated learning among the devices. The model parameters are transferred from each terminal device in the metaverse to the server more quickly thanks to the effectively developed network transmission technology. The combination of communication technology and FL will realize the comprehensive promotion of intelligence anytime and anywhere in the metaverse.

$\bullet$  \textbf{Applications}. Federated supervised learning algorithms can provide network environment analysis and intrusion prediction. A prediction model is a crucial tool for determining how to distribute wireless resources. Convex optimization, for instance, can be solved using the federated reinforcement learning technique. It breaks down the difficult issue into manageable components like resource management and network control. Therefore, federated learning can improve the intelligence of wireless networks, which is the goal pursued in the metaverse \cite{qin2021federated}. Unlike traditional machine learning, federated learning is able to avoid direct exposure of data to third parties at each terminal and has natural protection for data. The application of federated learning to wireless communication in the metaverse has great potential. For example, in \cite{samarakoon2019distributed}, federated learning is used for the estimation of key parameters in vehicular communication. Experiments demonstrate the scheme significantly preserves privacy while conserving transmission resources. 

\subsection{IoT of FL4M}

$\bullet$  \textbf{Purposes}. The Internet of Things (IoT) is an extended and expanded network based on the Internet that extends the interaction between users and objects. The IoT generates data through a large number of sensors and then uses artificial intelligence technology to accomplish various tasks. In the metaverse, the ways to link virtual and physical spaces are diverse. Therefore, IoT technology assists the metaverse in achieving data exchange and easy access.

$\bullet$  \textbf{Challenges}.  However, it is obvious that safely offloading the explosive data generated by each edge device to the server is a difficult problem. Furthermore, in the IoT, each user's data is heterogeneous \cite{nagar2019privacy}. It is unrealistic to apply traditional ML models to complex application scenarios in the metaverse.

$\bullet$  \textbf{Methods}. Recently, FL, a distributed collaborative AI approach, has been proposed to build an intelligent and privacy-enhanced IoT system \cite{nguyen2021federated}. The combination of FL and IoT technology in the metaverse not only provides privacy protection for distributed IoT systems but also reduces the communication delays caused by data offloading. In other words, in the metaverse, the combination of IoT and FL completes the transformation from physics to data while ensuring the privacy of users.

$\bullet$  \textbf{Applications}. Recent studies \cite{zeng2022hfedms, zhou2022resource} have shown that FL can be broadly applicable in the IoT industry, such as in smart healthcare and smart cities, etc.  In addition, FL also helps detect malicious attacks in federated IoT systems. The TensorFlow federated project (TFF) is an open-source federated learning platform developed by Google, and it contains two layers of interfaces: the federated layer API and the federated core (FC) API. Besides, synchronously optimized modeling algorithms like FedAvg \cite{mcmahan2017communication}, personalized federation learning schemes \cite{tan2022towards, wu2020personalized} in the metaverse are also applicable to build models adapted to different clients and scenarios. 

\subsection{Edge Computing of FL4M}

$\bullet$  \textbf{Purposes}. Edge computing \cite{shi2016edge} refers to computing generated near the device to provide low-latency services and more rapid response. In contrast, cloud computing \cite{dillon2010cloud} is processed by transferring data to a cloud computing center. Applications such as live video platforms and e-commerce platforms are inherently dependent on cloud computing. Each IoT device can select to transfer its intensive computing tasks to the edge devices via wireless connectivity, thereby reducing communication overhead. However, in some areas that require high real-time performance, cloud computing may have an incalculable impact on users due to high latency and network lag due to resource constraints. For example, if a self-driving car suffers a delay in its direct command during driving, failing the system to respond in a timely manner, it could lead to a traffic accident. Therefore, in the metaverse, a huge amount of data generated by massive IoT device connections requires more real-time and secure computing technology \cite{li2018learning}.

$\bullet$  \textbf{Challenges}. Edge computing meets this requirement, offering computation and storage to users close to edge devices while lowering the bandwidth pressure of data transmission to the cloud center server \cite{siriwardhana2021survey}. However, it still leaves the problem of user data being shared externally. As a result, some businesses are reluctant to share their private datasets with edge servers for fear that rivals will steal and use the information against them.

$\bullet$  \textbf{Methods}. FL was introduced into mobile edge computing (MEC) technology to ensure that the user's sample data is always stored locally on the device \cite{lim2020federated}. Collaboration among users requires only uploading the locally trained model parameters to the central server. A more desirable global model can be achieved with the collaboration of all participants.  In addition, even if a small number of edge devices are under attack or fail, it will not lead to problems of massive data privacy exposure or service interruption. Consequently, the integration of edge computing enhances the security and reliability of federated learning \cite{wan2022privacy}.

\begin{figure}[t]
	\centerline{\includegraphics[width=1\linewidth]{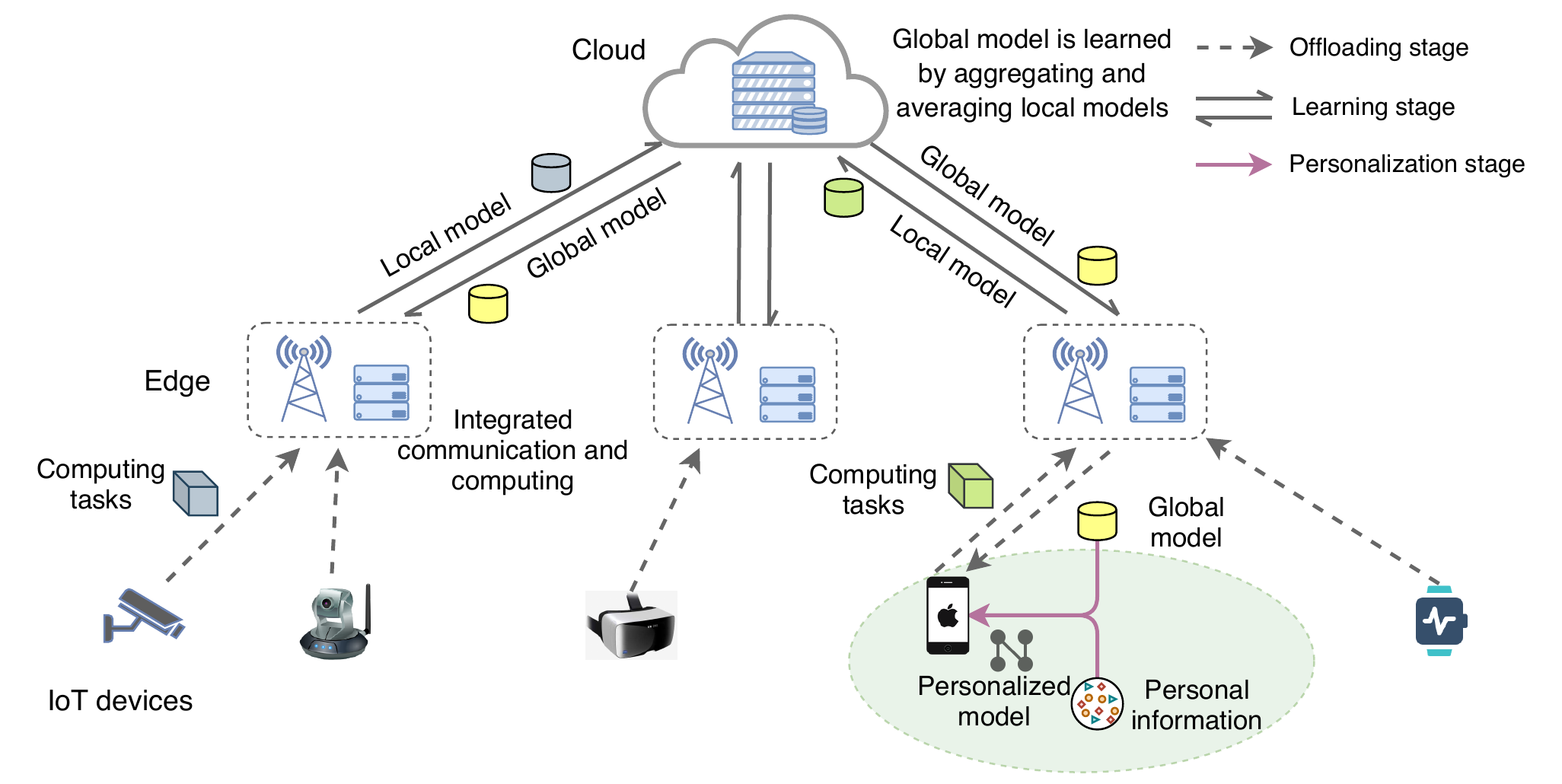}}
	\caption{Edge computing in FL for IoT applications \cite{wu2020personalized}.} \label{PerFit}
\end{figure}

$\bullet$  \textbf{Applications}. Figure \ref{PerFit} shows that the PerFit framework \cite{wu2020personalized} provides the necessary edge computing power for IoT devices. IoT devices are ubiquitous in the metaverse, which makes cyberattacks more varied and sophisticated. Combining edge computing with FL, network attack detection models guarantee users' privacy while collaborating with multiple devices for optimization.

\subsection{Blockchain of FL4M}

$\bullet$  \textbf{Purposes}. Different from other technologies used in the metaverse, blockchain not only establishes the connection between these technologies but also creates a complete closed loop with scenarios, behaviors, and actions. The metaverse constructs a distributed economic system based on two essential characteristics of blockchain, digital identity, and personal data rights confirmation.

$\bullet$  \textbf{Challenges}. The traditional centralized federated learning framework relies on the security of the centralized nodes, while one of the core requirements of blockchain in the metaverse is decentralization. Therefore, how to combine federated learning with blockchain is a challenging problem.

$\bullet$  \textbf{Methods}. At present, there have been many studies \cite{wang2021blockchain, baker2022blockchain, wan2022privacy} combining FL with blockchain, replacing the central aggregator with peer-to-peer blockchain, and utilizing blockchain nodes to aggregate the federated learning model. In this way, the security and privacy of data in distributed machine learning are guaranteed. Such distributed storage ensures the identity authentication of users and the security of digital assets. In particular, economic activities in the metaverse, such as real estate development and game rewards, are inseparable from cryptocurrencies and permanent records of digital transactions. Digital identities in the metaverse can be traced to some underlying data in such a transparent and open paradigm.

$\bullet$  \textbf{Applications}. Integrating blockchain and FL can prevent model poisoning attacks and form a more stealthy defense mechanism \cite{unal2021integration}. Kang \textit{et al.} \cite{kang2022blockchain} proposed the application of combining blockchain and FL in the industrial metaverse. By establishing a cross-chain authorized federated learning framework, it is possible to conduct safe and protected data training in virtual space and real space. In the metaverse, blockchain and federated learning can complement each other. Blockchain facilitates federated learning to establish a learning incentive mechanism and realize the automatic distribution of benefits.

\subsection{Extended Reality of FL4M}

$\bullet$  \textbf{Purposes}. Extended Reality (XR) is essential in the metaverse to achieve the interaction between the virtual and real worlds. XR is widely considered to be the key to opening the door to the metaverse. Extended reality includes VR (virtual reality), AR (augmented reality), and MR (mixed reality). VR employs devices to simulate a virtual environment to assist users in fully immersing themselves in the virtual world. On the other hand, AR draws virtual images in the real world by using imaging technology, allowing users to experience the intersection of the real and virtual worlds. MR incorporates the technical features of VR and AR.

$\bullet$  \textbf{Challenges}. Of course, hardware equipment like wearable headsets or smart terminals is necessary for the technological realization of XR. The scene construction of these devices is inseparable from the recognition and target-tracking algorithms. However, those target detection algorithms require a lot of data training to obtain an accurate model while only a limited amount of data can be stored on mobile devices.

$\bullet$  \textbf{Methods}. An illustration of XR of FL4M is shown in Figure \ref{FLandXR}. Users in the real world interact with the virtual world through XR technology. Each user has an avatar and a large amount of data in the metaverse. These users can perform federated learning in the metaverse. That is, these users use their own devices and data to jointly participate in a machine learning task.

$\bullet$  \textbf{Applications}. The study \cite{zhou2022mobile} applied FL with mobile augmented reality to address the issues of data storage and model accuracy. It also discusses the feasibility of the FL-MAR system and proposes an optimized resource allocation scheme for the FL-MAR system. As a result, the metaverse leverages the combination of XR and FL more effectively in fields such as education, shopping, and autonomous driving. Everyone has varied shopping preferences and lifestyles, for instance. Combining FL and XR provides customers with better virtual shopping experiences and more accurate shopping recommendations by meeting the local training models of different mobile devices.

\section{Key Challenges} \label{challenges}

\begin{figure}[h]
    \centerline{\includegraphics[width=1\linewidth]{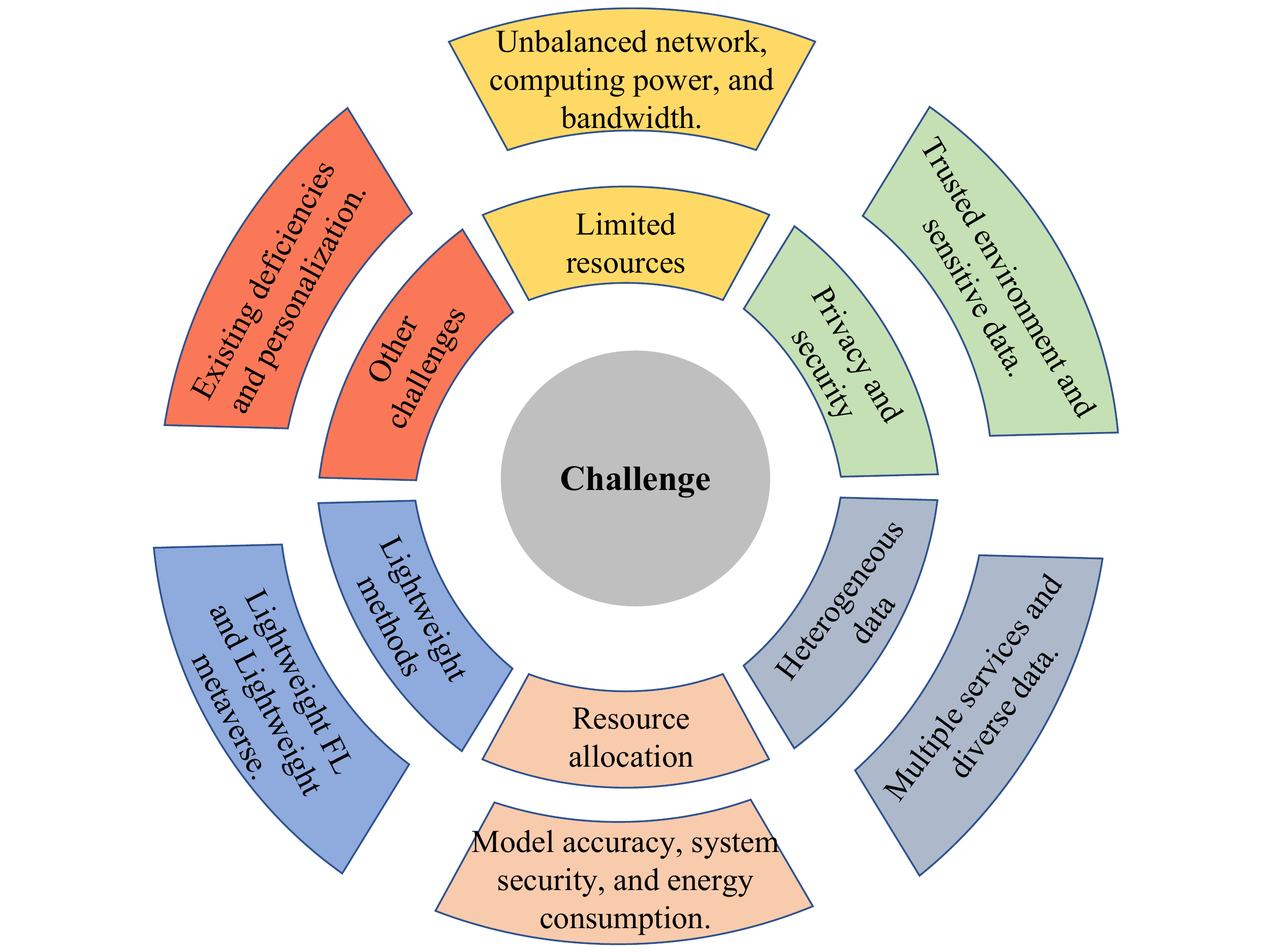}}
    \caption{The keywords of each challenge.} 
    \label{FivePart}
\end{figure}

Federated learning and the metaverse are two promising concepts. The former can address the issue of security in model transmission and sharing to an extent. The metaverse aims to improve the interactivity between the physical and virtual worlds, as well as to improve the user experience. The convergence of their uses is not a simple combination of technological solutions. That is, it is not just AI research with privacy protection in various areas of the metaverse. Therefore, in our opinion, the following challenging issues are worth exploring and discussing in detail. Figure \ref{FivePart} lists the keywords of each challenge to give an overview of the FL4M challenges. Details are described below.

$\bullet$ \textbf{Limited resources}. For a system or framework, resources are often limited and need to be scheduled and allocated appropriately. The problem of limited resources may be caused by various reasons, such as the heterogeneity of data. For applications and deployments with multiple unbalanced network nodes and metaverse use, a lot of content is difficult to accomplish without a cluster of computing networks \cite{imteaj2021survey}. When there is a shortage of some important computing resources, the whole system may quickly reach a bottleneck or fail to complete the task. Some problems involving communication and computation also need to be solved. Meanwhile, a large amount of data would be generated. The data may be valuable but difficult to organize. This may make the entire FL process more burdensome. The first issue is how to handle the data generated by different devices with limited resources. Another fundamental issue is the trade-off between data utility and system efficiency. With fewer data available, there is a need to ensure that the whole system operates efficiently and accurately. For the accuracy of the model, this may cause additional overhead. In addition, interactions at the metaverse level need to be taken into account. In a real-world society, it is not reasonable to make users perform long waits or complex behaviors. The design of the entire application requires the integration of many process details.

$\bullet$  \textbf{Privacy and security}. In the applications of the metaverse, users often generate a variety of large amounts of data in their activities. At the same time, various devices that connect the virtual and real worlds also collect data related to users. This raises the issue of users' privacy protection. In an untrustworthy metaverse environment, this may lead to a waste of resources or deterioration of model performance in the training of FL. Although privacy protection is a key concern of FL techniques, there is still the risk of private data leakage in the metaverse. The data in the collection of devices, the interaction between the real and virtual worlds, and the transmission of the FL framework may have the possibility of leakage. The metaverse is the end closer to the data and requires greater responsibility for the supervision and storage of the data. Once data is at risk here, it makes no sense, even if FL strictly protects it. Besides, handling large data in the metaverse for FL requires an efficient and lightweight security and privacy solution. Security and efficiency need to be considered simultaneously. Some security mechanisms, such as the RSA encryption method, while reliable enough, are also not applicable to scenarios with large data. FL needs to focus on what perspective and in what way to protect the data. For example, the random response approach in FL is efficient, but it is not always able to achieve the complex data generated by the devices of the metaverse.

$\bullet$  \textbf{Heterogeneous data}. To maintain data utility, the metaverse must handle multi-source heterogeneous data. The quality of metaverse's work is influenced by the source and completeness of the data. Especially in predictive applications, if the data is collected by inaccurate devices, the overall performance may be poor. The problem of data heterogeneity in FL is mainly caused by the diverse data of each client involved in the training. These data are non-independent and identically distributed (non-IID), often leading to severe degradation of model accuracy. Effective and comprehensive segmentation strategies and datasets can cover different non-IID scenarios. There is a need to devise efficient methods for FL to address the redundancy and unrealistic nature of heterogeneous data in the metaverse. In the industrial metaverse, HFedMS \cite{zeng2022hfedms} reduces the data heterogeneity of FL by using dynamic grouping and training mode conversion. Historical data semantics are compressed to compensate for forgotten knowledge. However, how can we solve this challenge in other scenarios? It is worth noting that the multiple devices of the meta-universe also introduce other aspects of heterogeneity. It is difficult to ensure that the protocols, formats, and workflows are consistent across devices. Although complex settings are freer, they also add a certain degree of burden to a system.

$\bullet$ \textbf{Resource allocation}. There is a crucial challenge in allocating resources in a rational and effective way. The limited resources make it necessary for FL to run in a better-planned way in the metaverse to avoid some extreme cases, such as too much running time and poorly trained models. Multiple metrics, including total energy consumption, system security, and model accuracy, form a weighted optimization problem. Low bandwidth makes the training of FL models or the response of metaverse applications too slow. Devices with different computing powers also cause the overall performance of the training to be uneven or users to become biased. It is difficult to consider how to design FL solutions that can run stably with low or unbalanced computing resources according to the scenario requirements. Some applications seek greater system security, while others seek better model accuracy. Although there are some algorithms and schemes \cite{wangyitong2022survey, zhou2022resource} proposed to solve the load and resource allocation problems of the metaverse systems, they do not take into account the dynamic cases. The number of system nodes is usually not determined at the time of application deployment, and the allocation of resources should change dynamically as nodes are added or removed. In addition, the robustness of the resource allocation scheme needs to be considered in order to cope with extreme situations.

$\bullet$ \textbf{Lightweight methods}. Metaverse applications can be accessed and interacted with at any time through IoT devices. The highly immersive virtual experience provided to users is served by a large amount of human-computer interaction (HCI) \cite{karray2008human}. The large number of devices in FL also brings expensive computation costs. With limited computing and communication resources, some traditional large-scale deployment schemes are not suitable for IoT devices. The design of lightweight models not only simplifies the process of accessing applications of the metaverse but also reduces the burden of training in FL. Existing metaverse applications still lack lightweight AI techniques. The relevant definitions are not very clear in the industry. MetaAID \cite{zhu2022metaaid} is a flexible framework designed to support the language and semantic technologies in digital twins and virtual humans. There are also many lightweight federated schemes. A lightweight privacy-preserving scheme of FL is proposed by Wei \textit{et al.} \cite{wei2021lightweight} to protect the security of all local data. To deal with large-scale FL tasks, a secure mask-reusing mechanism is also designed. In the research of federated recommender systems, Zhang \textit{et al.} \cite{zhang2022lightfr} proposed a model called LightFR that is able to perform tasks well in the condition of devices with limited computing resources. 

$\bullet$ \textbf{Other challenges}. In addition to the challenges mentioned above, there are others that deserve the attention of researchers. For example, the personalization of FL in metaverse applications is an act that allows specific services and content empowerment for different users. One of the emerging challenges in the metaverse is how to perform personalized model learning or personalized solutions. To prevent the overall system from failing, the issue of FL attacks and defenses should also be addressed in the metaverse. Besides, some of the deficiencies of FL can also be solved or minimized in future studies using metaverse techniques. It will welcome both practical and theoretical contributions for solutions to the interpretability/explainability problems in FL4M.

\section{Promising Directions} \label{directions}

In our opinion, integrating FL into the metaverse is conducive to building an open, fair, and safe metaverse. At present, FL4M is still in its initial stage, and there are many research problems. Therefore, in this section, we discuss several potential research directions of FL4M, including the incentive mechanism of FL4M, intellectual property right protection in FL4M, the model fairness of FL4M, the privacy-utility trade-off in FL4M, catastrophic forgetting in FL4M, and the application of FL4M.

$\bullet$ \textbf{Incentive mechanism of FL4M}. In the future, people will also live in the metaverse when they are in the virtual world. Everyone's information and activity data are recorded as data. Therefore, each user in the metaverse has a large number of digital assets. However, we cannot ignore two important issues. How to encourage people to contribute data to the construction of the metaverse? How do we encourage people to use their own data and equipment to participate in machine learning tasks in the metaverse? A reasonable incentive mechanism is a solution to these problems. The design goal of the incentive mechanism is to distribute rewards fairly according to everyone's contribution, so as to encourage more people to join the machine learning tasks in the metaverse. A reasonable incentive mechanism can increase the enthusiasm of data asset owners. Furthermore, the incentive mechanism has the potential to reduce the behavior of malicious users who intentionally send incorrect data. Therefore, designing a reasonable incentive mechanism in the field of FL4M is a promising research direction. We believe that a reasonable incentive mechanism has at least several characteristics, such as trusted, fairness, credibility, accuracy, and timeliness.

$\bullet$ \textbf{Intellectual property right protection in FL4M}. In the field of FL4M, the protection of model intellectual property rights is an important problem. Training a federated learning model successfully usually requires a lot of resources. The computational and storage resources of each participant are consumed. The trained model should be considered the product of the labor of all participants. Therefore, participants should own the corresponding intellectual property rights. The protection of model intellectual property rights should not only ensure that the participants obtain the corresponding model intellectual property rights but also protect the model from illegal copying, tampering, redistribution, and abuse. Some hackers may steal the model through illegal means and then tamper with the model or claim ownership of the model. These behaviors will damage the interests and intellectual property rights of model owners. In this case, verifying the ownership of the model and thus protecting the model owner's intellectual property is a matter of great concern. The issue of model intellectual property protection is an interdisciplinary and comprehensive topic involving computer security, intellectual property protection, law, and many other aspects.

$\bullet$ \textbf{Model fairness of FL4M}. FL4M is easy to have the problem of model unfairness. Model unfairness occurs in model training. The distribution of training data or the deviation in the model makes the model perform unevenly on some sensitive features. A common phenomenon is that individual feature data are underrepresented in the total dataset, so the weights obtained after the model has been trained are not representative of the problem. For example, we use gender as a sensitive feature. If the ratio of male to female participants is 100:1, then samples containing males will dominate the model training phase. Thereby, the model may not learn accurately for the female sample. Unbalanced data distribution is a very common phenomenon in the FL4M field. Since FL training models require multiple participants, each participant is a data owner. The participants have different characteristics and behaviors. The data distribution among participants may be different in different regions. Therefore, in the field of FL4M, it is easy to have model unfairness due to unbalanced data distribution. Model unfairness can seriously damage the accuracy of the model. How to introduce fairness into the model is an important research topic. This allows the model to produce fair output even if it is trained on unfair data. This research direction is very important because unbalanced data distribution is common.

$\bullet$ \textbf{Privacy-utility trade-off in FL4M}. One of the core issues of FL4M is how to achieve a balance between privacy and utility. Privacy can be quantified through information privacy \cite{zhang2022no}. Accuracy can be used to quantify utility. Quantifying the privacy utility, however, is a difficult problem. We are aware that security and availability are opposing factors. Based on practical requirements, we should consider security and availability together in order to maximize availability while meeting security requirements. This frequently necessitates a great deal of knowledge and effort. As a result, we wonder if there is a unified standard or framework for calculating privacy utility. Furthermore, whether there are algorithms that optimize security, model performance, and accuracy all at the same time is an important research question. Privacy and utility are frequently required in practical application scenarios. There is no doubt that the FL4M algorithms should strike a balance between privacy and utility.

$\bullet$ \textbf{Catastrophic forgetting in FL4M}. Catastrophic forgetting refers to an artificial intelligence system (such as a deep learning model) forgetting or losing some of its previously acquired abilities while learning a new task or adapting to a new environment. In the FL4M domain, the following scenarios can cause or exacerbate the phenomenon of catastrophic forgetting. i) When data are not distributed independently and identically across different clients, the model may forget previous knowledge gained by training with other clients' data. This phenomenon arises because of the discrepancy between the local and global data distributions. ii) In the federated category incremental learning problem, participants may often collect new categories. Since each participant's device has very limited storage space, it is difficult to keep a sufficient amount of data for all the categories. In this case, the performance of the model on the old category data is likely to suffer from severe catastrophic forgetting. iii) When new participants are added during the model training, as these new participants often have a large amount of new data. The catastrophic forgetting of the global model can be further exacerbated in this case. Catastrophic forgetting leads to a higher rate of loss of training data and a longer time to train the model. Therefore, how to solve the problem of catastrophic forgetting in FL4M is a promising research direction.

$\bullet$ \textbf{Application for FL4M}. Finding FL4M application scenarios and using FL4M to solve problems is an important direction for research. Until now, there have been a lot of studies that focus on how the metaverse can be used in situations like education, games, healthcare, social networks, advertising, marketing, etc. In particular, some literature discusses the application scenarios of FL4M in detail. For example, Kang \textit{et al.} \cite{kang2022blockchain} used FL to analyze sensing data in the industrial metaverse. In addition, some studies focus on the combination of FL and other technologies in the metaverse to solve practical problems. For example, Zhou \textit{et al.} \cite{zhou2022mobile} discussed the necessity and rationality of the combination of FL and MAR in the metaverse and studied two cases of the metaverse FL-MAR system. When developing technology, we must consider not only the technology itself but also the technology's application scenarios. When technology is thought to add significant value to certain fields, it will naturally advance. As a result, it is critical to investigate FL4M application scenarios (e.g., FL4M for healthcare, edge devices, advertising, social networks, blockchain, web search, etc.). The investigation of FL4M application scenarios may attract additional research and expose the shortcomings of the technologies when FL4M is used.

\section{Conclusion}  \label{sec:conclusion}

The metaverse, which is still in its early stages, is facing data challenges that will stymie its development. With increasing user privacy protection awareness and competition among enterprises or institutions, data acquisition becomes increasingly difficult. Data acquisition can be made easier and more legal in a FL-enabled metaverse. In this paper, we provide an introduction to FL4M and examine the necessity and rationality of studying FL4M in depth. FL4M not only protects users' data privacy to some extent, but it also makes use of each user's data, computational power, and model-building abilities. Further research on FL4M contributes to the creation of a future metaverse that is open, fair, and secure. The combination of some key metaverse and FL technologies is discussed in this paper. Finally, we investigate FL4M's challenges and promising directions. Remember that FL4M is likely to receive more algorithmic and theoretical improvements in the future. We hope that this article will stimulate the interest of more researchers in FL4M and inspire them to conduct more innovative research.

\begin{acks}
    This research was supported in part by the National Natural Science Foundation of China (Nos. 62002136, 62272196, and 61932011), Fundamental Research Funds for the Central Universities of Jinan University (No. 21622416), Guangzhou Basic and Applied Basic Research Foundation (No. 202102020277), Natural Science Foundation of Guangdong Province (Nos. 2022A1515011861 and 2019B1515120010), Guangdong Key R\&D Plan2020 (No. 2020B0101090002), National Key R\&D Plan of China (No. 2020YFB1005600), the Young Scholar Program of Pazhou Lab (No. PZL2021KF0023), National Joint Engineering Research Center for Network Security Detection and Protection Technology, and Guangdong Key Laboratory for Data Security and Privacy Preserving. Dr. Wensheng Gan is the corresponding author of this paper. 
\end{acks}

\pagebreak
\bibliographystyle{ACM-Reference-Format}
\bibliography{paper}

\appendix
\renewcommand\thefigure{\Alph{section}\arabic{figure}}    
\section{Picture appendix}
\setcounter{figure}{0}    
\begin{figure}[H]
	\centerline{\includegraphics[width=0.88\linewidth]{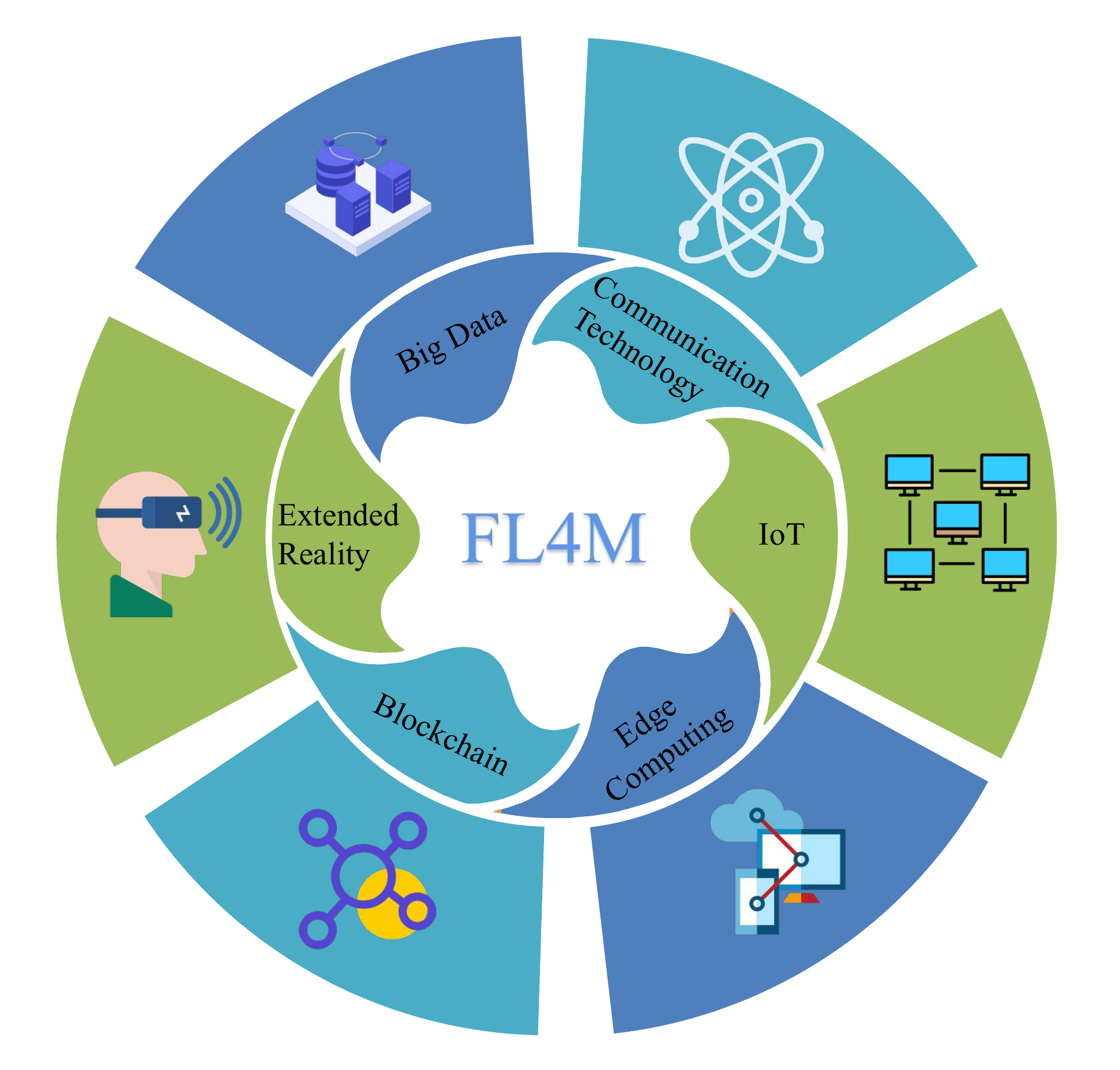}}
	\caption{Six key techniques in FL4M.} 
    \label{Key technologies}
\end{figure}

\setcounter{figure}{1}
\begin{figure}[H]
	\centerline{\includegraphics[width=0.9\linewidth]{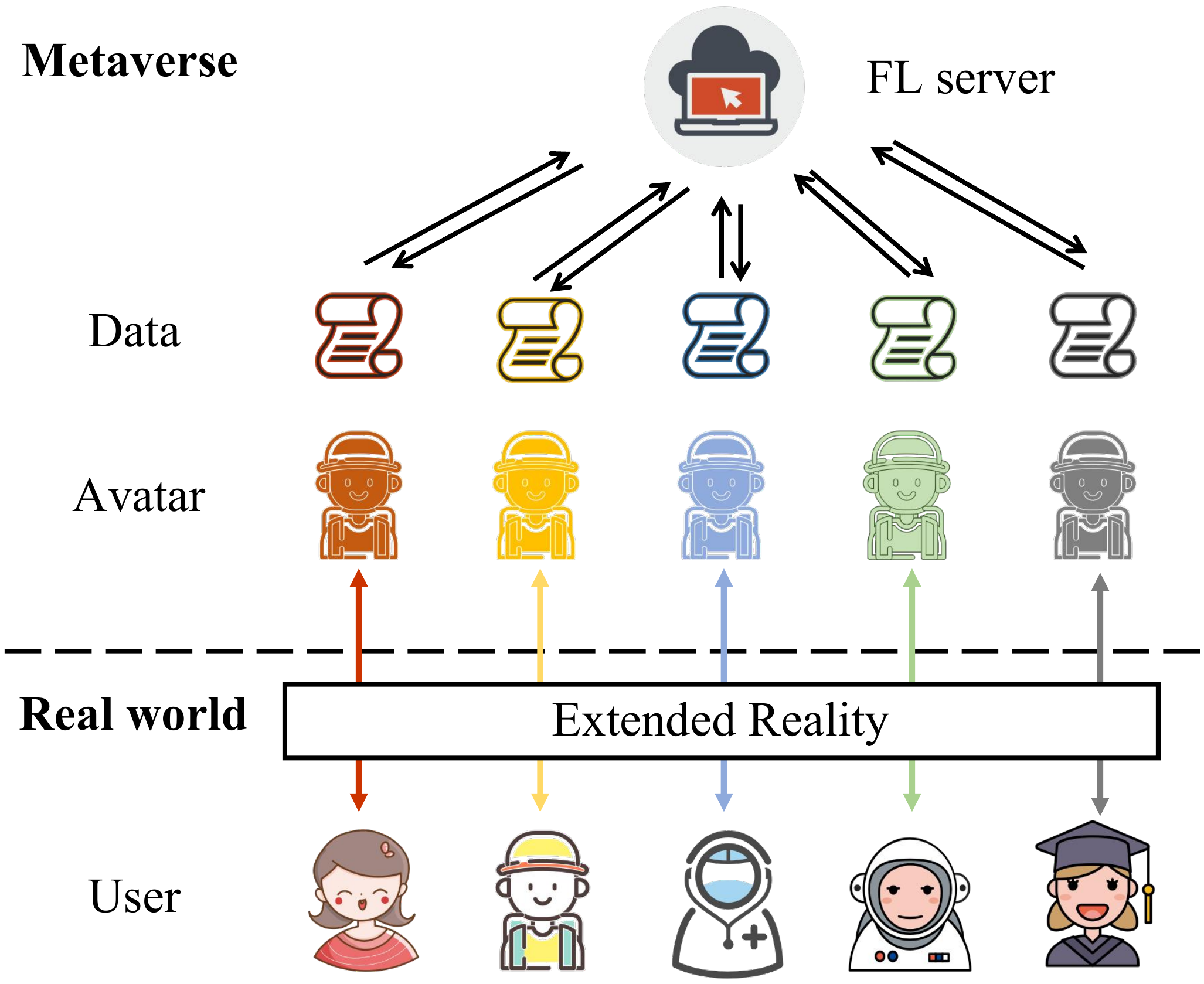}}
	\caption{Illustration of extended reality of FL4M.} 
    \label{FLandXR}
\end{figure}

\end{document}